\documentclass[useAMS,usenatbib]{mn2e}
\usepackage[dvips]{graphicx}

\title[The Chemical composition of the post-AGB F-supergiant CRL 2688]{The Chemical composition of the post-AGB F-supergiant CRL 2688 }
\author[Miho~N. Ishigaki et al.]{Miho~N. Ishigaki$^{1}$
\thanks{E-mail:ishigaki.miho@nao.ac.jp}
, Mudumba Parthasarathy$^{1,2}$, Bacham~E. Reddy$^{3}$, \newauthor
Pedro Garc\'{i}a-Lario$^{4}$, Yoichi Takeda$^{1}$, Wako Aoki$^{1}$, \newauthor
D.~A. Garc\'{i}a-Hern\'{a}ndez$^{5,6}$, and Arturo Manchado$^{5,6}$\\
$^{1}$National Astronomical Observatory of Japan, 2-21-1 Osawa, Mitaka, Tokyo 181-8588, Japan\\ 
$^{2}$Aryabhatta Research Institute of Observational Sciences, Nainital, India \\
$^{3}$Indian Institute of Astrophysics, Bangalore 560034, India \\
$^{4}$Herschel Science Centre, European Space Astronomy Centre, Villafranca del Castillo, P.O. Box 78, E-28080 Madrid, Spain \\
$^{5}$Instituto de Astrof\'{i}sica de Canarias, 38200, La Laguna, Tenerife, Spain \\
$^{6}$Departamento de Astrof\'{i}sica, Universidad de La Laguna, 38200, La Laguna, Tenerife, Spain
}

\begin{document}

\date{}

\pagerange{\pageref{firstpage}--\pageref{lastpage}} \pubyear{2012}

\maketitle

\label{firstpage}

\begin{abstract}

We present an analysis of a high resolution ($R\sim 50,000$) optical
spectrum of the central region of the proto-planetary nebula CRL 2688.
 This object is thought to have recently moved off the AGB, 
and display abundance patterns of CNO and heavy elements 
that can provide us with important clues to understand the 
nucleosynthesis, dredge-up and mixing 
experienced by the envelope of the central star during its AGB stage of 
evolution. The analysis 
of the molecular features, presumably originated from the circumstellar 
matter provides further constraints on the 
chemistry and velocity of the expanding shell, expelled as a 
consequence of the strong mass loss experienced by the central star.

We confirm that the central star shows a spectrum typical of an 
F-type supergiant 
with $T_{\rm eff}=7250\pm 400$ K, $\log g=0.5$ and [Fe/H]$=-0.3\pm 0.1$ dex. 
We find that the 
abundance pattern of this object is characterized by enhancements of 
Carbon ([C/Fe]$=0.6\pm 0.1$), Nitrogen ([N/Fe]$=1.0\pm 0.3$) 
and Na ([Na/Fe]$=0.7\pm 0.1$), similar to 
other previously known carbon-rich post-AGB stars. Yttrium 
is also enhanced while the [Ba/Y] ratio is 
very low ($-1.0$), indicating that only the light s-process elements are 
enhanced. The Zinc abundance is found to be normal, [Zn/Fe]$=0.0\pm 0.3$, 
suggesting that there is no depletion of 
refractory elements. 
The \mbox{H\,$\alpha$}, \mbox{Na\,{\sc i}} and \mbox{K\,{\sc i}} resonance 
lines show prominent emission components, 
whose helio-centric radial velocities are offsetted by $-41\pm 3$ km s$^{-1}$ 
relative to the photospheric metal-absorption lines. 
The molecular features of C$_{2}$ and CN also show emission components, 
whose velocities are consistent with the emission components of the 
\mbox{H\,$\alpha$}, \mbox{Na\,{\sc i}} 
and \mbox{K\,{\sc i}} lines. 
On the other hand, their absorption components are more highly blue
shifted than the corresponding emission components, which suggests that 
the regions where the emission and absorption components arise 
are expanding at different velocities.  
   
\end{abstract}


\section{Introduction}

 CRL 2688 (V1610 Cyg, Egg Nebula) is a  bipolar proto-planetary nebula (PPN).
 The central star is thought to be a F5Iae post-AGB supergiant. 
Initial studies of CRL 2688
were mostly based on low resolution spectrophotometric and spectropolarimetric
 data and were made by \citet{ney75}, \citet{forrest75}, \citet{crampton75} and 
\citet{cohen77,cohen80}. The low resolution spectrophotometric observations
revealed emission from C$_2$, C$_3$ and SiC$_2$. The Swan bands of C$_2$ are also seen 
in emission from the bipolar lobes \citep{cohen80}. 
Because the object is faint ($V = 13.5$)
with significant circumstellar reddening its high resolution 
spectroscopy was not
carried out until the year 2000. The first high resolution ($R=15,000$) 
optical spectra were 
analyzed by \citet{klochkova00}. They found that the central star is a F5Iae
post-AGB supergiant with overabundance of carbon and s-process elements. 
\citet{klochkova00} derived $T_{\rm eff}=6500$ K, $\log g = 0.0$, microturbulent velocity $= 6.0$ km s$^{-1}$
and [Fe/H]$= -0.59$. 

High-spatial resolution {\it Hubble Space Telescope} imaging 
revealed the detailed morphology of this object; a flattened dust cocoon 
obscuring the central star as well as extended bipolar nebulae exhibiting 
numerous concentric arcs crossed by a pair of beam-like structures at the both 
sides of the equatorial plane \citep{sahai98}. These arcs 
are suggested to be associated with repeated mass-loss events
during the AGB phase \citep{sahai98,balick12}.  

Recently,  \citet{wesson10}
studied the {\it Herschel Space Observatory}
- SPIRE FTS spectra of CRL 2688 covering the wavelength
range 195 to 670 $\mu$m. They found the far infrared spectrum
of CRL 2688 to be very complex with about 18 different species of complex
molecules. They also detected water in CRL 2688 for the first time.

CRL 2688  shows small amplitude light variations
with a period of about 90 days \citep{hrivnak10} and  
some of the absorption
and emission lines seems to show a radial velocity 
variability with time  
\citep{klochkova00}. 
The variability indicates that the central star 
as well as the circumstellar material is undergoing 
rapid variations in their dynamical and chemical 
properties possibly due to pulsation and mass-loss. 

Mechanisms that give rise to observed complex morphology, chemistry 
and variability of CRL 2688 as well as other post-AGB stars
 are still far from being completely understood.
Identifying constraints on these mechanisms is important 
for understanding the chemical evolution of our Galaxy since 
low-to-intermediate mass stars on the AGB phase of evolution are 
thought to be mainly responsible for synthesizing 
and contributing with CNO and s-process elements to the interstellar 
medium via mass loss.

In this paper we present an analysis of a much higher resolution 
spectrum of CRL 2688.  \\

\section{Observation and reduction}

A high resolution ($R\sim 50,000$) spectrum of CRL 2688 was obtained
with the Utrecht Echelle Spectrograph (UES) 
at the 4.2 m William Herschel Telescope on July 14, 2001. 
The slit of 1 arcsec widths was positioned at the central star region.
The wavelength range from 4300 to 9000 {\AA} was observed 
with an exposure time of 1200 sec. 
The data were reduced with standard IRAF routines. 
The signal to noise ratio of the reduced spectrum per resolution 
element varies from 30 to 70 in this wavelength range. 

Equivalent widths (EW) of atomic absorption lines were measured 
by fitting Gaussian to each feature using a linelist described below.

\section{Description of the spectrum}

\begin{figure*}
\centering
\begin{minipage}{128mm}
\includegraphics[width=12cm]{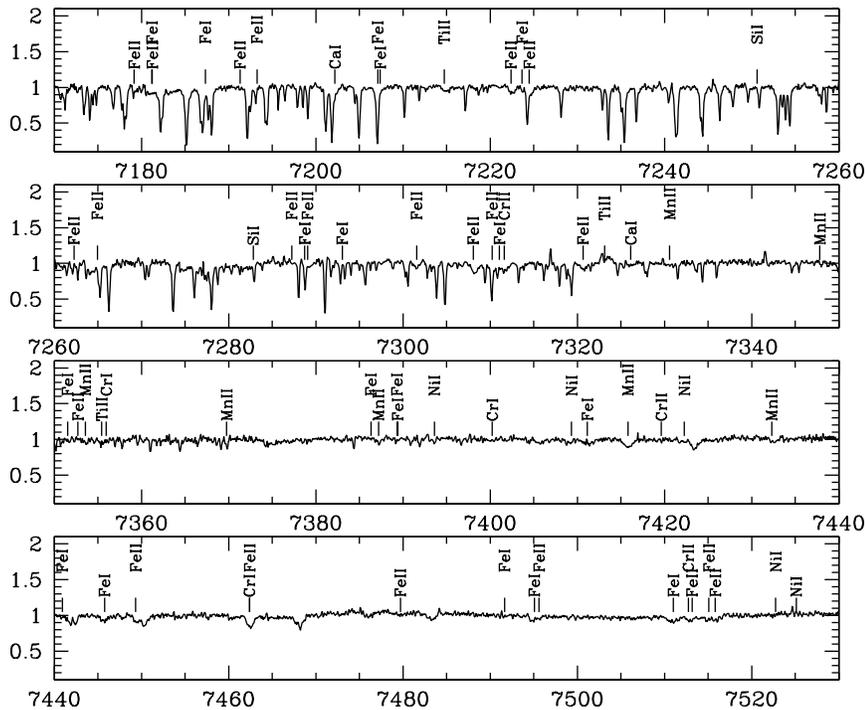}
\end{minipage}
\vspace{-2.0cm}
\caption{A portion of the obtained spectrum of CRL2688. 
Positions of 
metal-lines used in the EW determination are indicated. 
}
\label{fig:spec_example}
\end{figure*}

The reduced spectrum shows various photospheric absorption lines and molecular 
lines. A portion of the observed spectrum is shown in 
Figure \ref{fig:spec_example}. Absorption lines that are 
considered in the EW measurements taken from \citet{reyniers04} and 
\citet{klochkova00} are marked.

\subsection{Atomic lines}

\begin{figure}[htbp]
\begin{center}
\includegraphics[width=8cm]{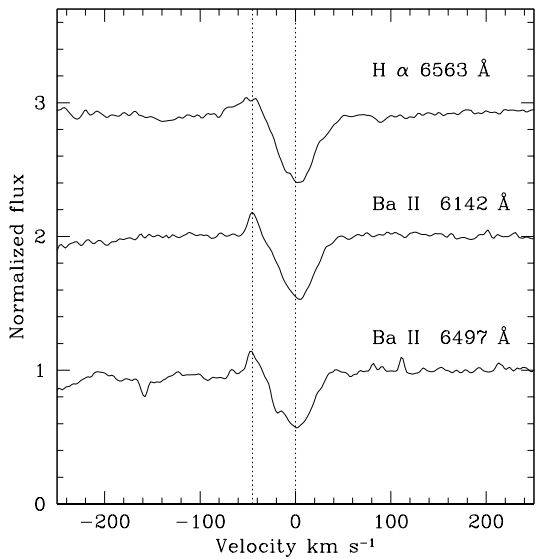}
\caption{Line profile for \mbox{H\,$\alpha$} and \mbox{Ba\, {\sc ii}} lines. Only the core region 
of the line profile is 
shown for the H$\alpha$ line. All lines are shifted in velocity 
relative to the photospheric one.}
\label{fig:pcyg}
\end{center}
\end{figure}

The obtained optical spectrum of CRL 2688 exhibits 
numerous atomic absorption lines of metals, most of which 
are too weak to be used in the abundance analysis, as illustrated in 
Figure \ref{fig:spec_example}.   
Many of the relatively strong lines like \mbox{H\,$\alpha$} and \mbox{Ba\,{\sc ii}} 
show a blueshifted emission 
component superimposed on the absorption component 
as shown in Figure \ref{fig:pcyg}.
The presence of emission component in the \mbox{H\,$\alpha$} 
violet wing is in agreement with the suggestion from a 
spectropolarimetric observation of \citet{klochkova04}.   
The observed features indicate mass loss from the central star.

\subsection{Molecular lines}
\label{sec:mols}

\begin{figure*}
\begin{minipage}{128mm}
\includegraphics[width=12cm]{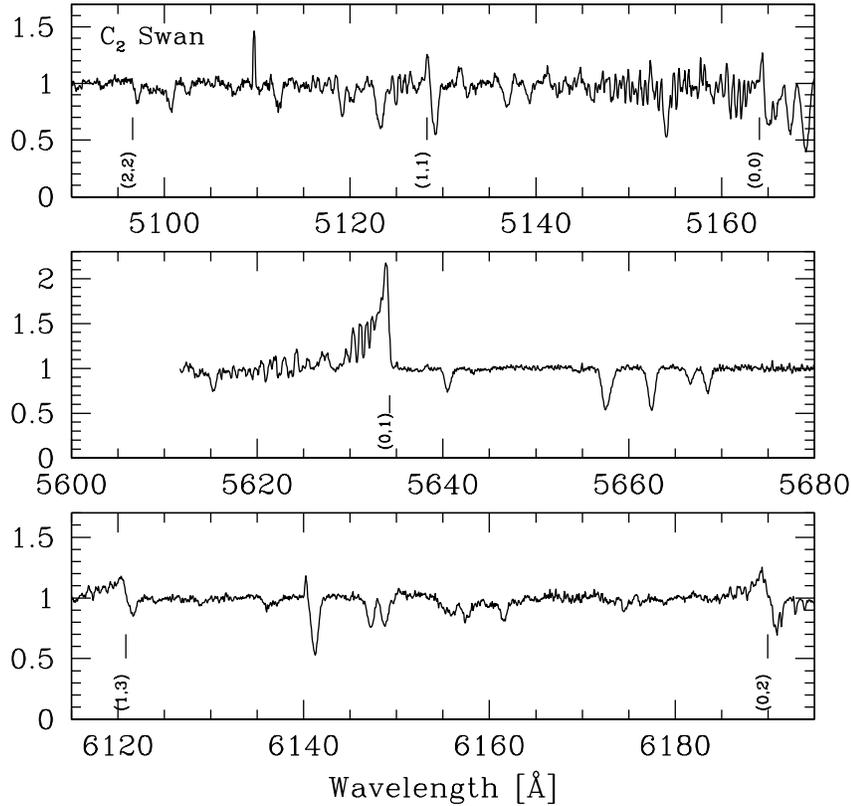}
\end{minipage}
\caption{The observed spectrum of the C$_{2}$ Swan system. 
The C$_2$ emission features are indicated based on the identification
by \citet{cohen77}.}
\label{fig:cmol1}
\end{figure*}

\begin{figure*}
\begin{tabular}{c}
\begin{minipage}{128mm}
\includegraphics[width=11.5cm]{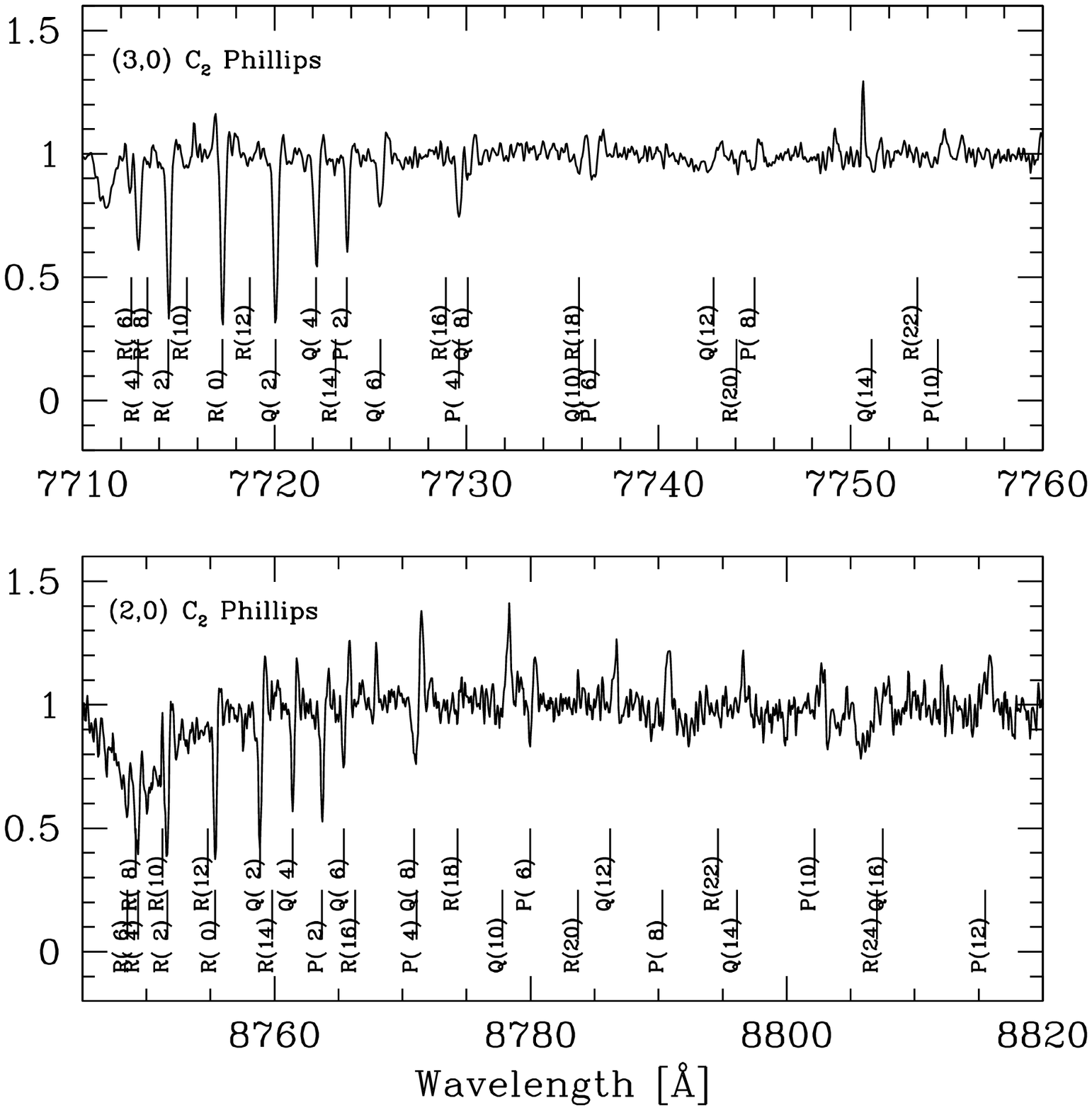}
\end{minipage}\\
\begin{minipage}{128mm}
\includegraphics[width=11.5cm]{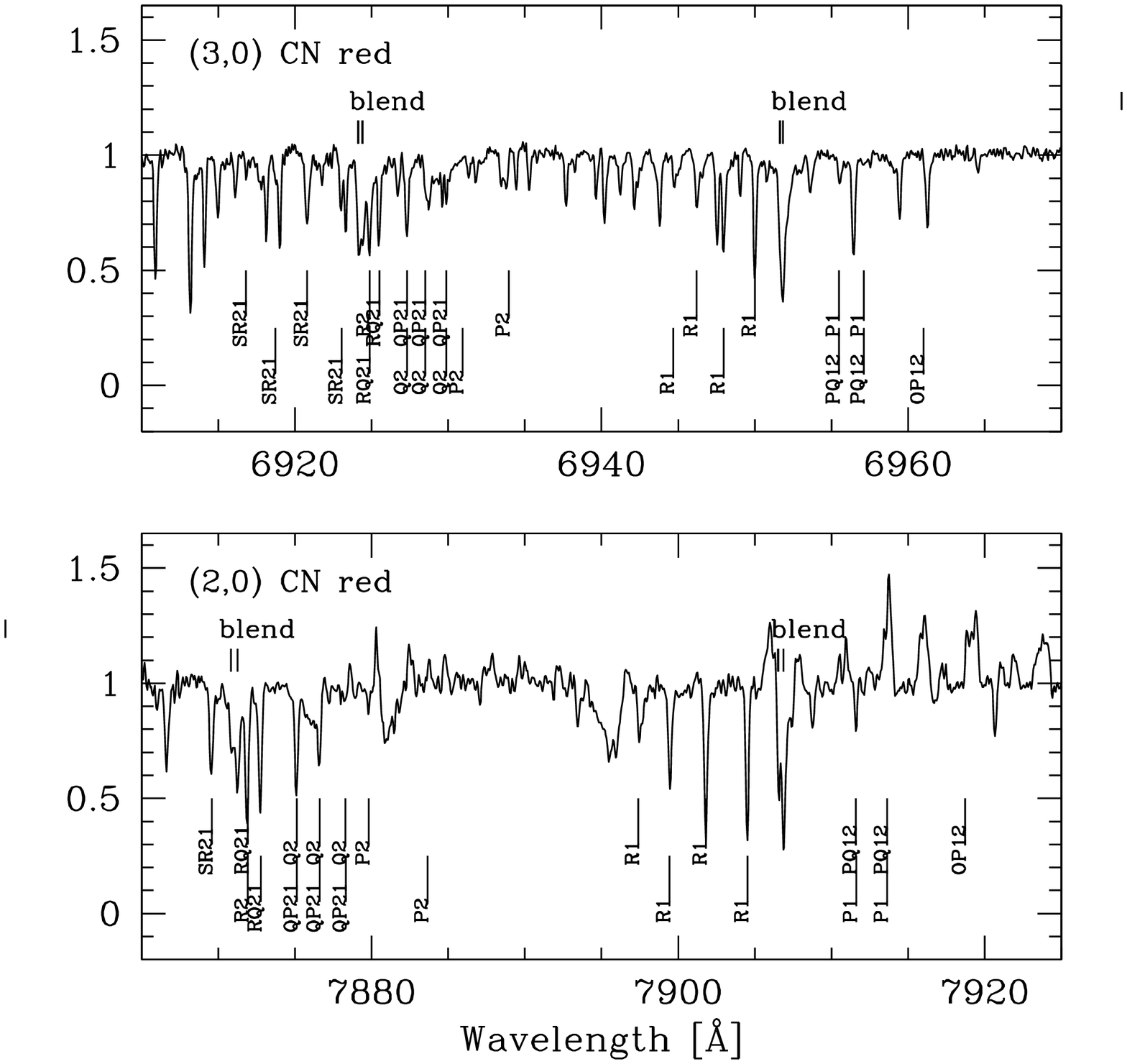}
\end{minipage}
\end{tabular}
\caption{The observed spectrum of the C$_{2}$ Phillips system (top) and 
the CN red system (bottom).  
Identification of molecular features are made based on the linelist of 
\citet{bakker96}. The blended features are 
indicated as 'blend'.}
\label{fig:cmol2}
\end{figure*}

 The molecular lines in this star are likely originated from its
circumstellar envelope, because such molecules would be 
destroyed in its photosphere with $T_{\rm eff} \sim 7000$ K. 
The spectral regions dominated by molecular bands are 
shown in Figure \ref{fig:cmol1} and \ref{fig:cmol2}.

Strong emission bands of the C$_{2}$ Swan system are identified 
consistent with previous studies \citep{crampton75,cohen80,klochkova00,klochkova04}. 
The particularly strong band at 5635.5 {\AA} of C$_{2}$(0,1)
is shown in the middle panel of Figure \ref{fig:cmol1}. \citet{cohen77} 
showed that the C$_{2}$ emission bands are prominently seen only 
in unpolarized light. The negligible polarization for the 
C$_{2}$ Swan emission bands has also reported 
by the study of \citet{klochkova04} 
with a much higher-resolution spectropolarimetry, 
 suggesting that the observed C$_{2}$ emission 
may come directly from the line forming region.

We confirm the presence of molecular absorption bands of the 
C$_{2}$ Phillips system and the 
CN red system, as shown in Figure \ref{fig:cmol2}.
The detection of these molecular absorption bands 
had previously been reported by \citet{bakker97}. 
Additionally, some of the features in 
C$_{2}$ Phillips (2,0) and (3,0) bands are shown in emission, whose 
central wavelength is less blue shifted than the 
corresponding absorption component. A few transitions 
in the CN (2,0) band also appear to show emission components.

\citet{cohen80} have suggested the presence of 
the Merrill-Sanford bands of SiC$_{2}$ in absorption, strongest  
at 4977 {\AA}, particularly in the region of the nebular lobe. 
Our spectrum does not show this feature in agreement with 
the apparent lack of this feature in the study of the 
central region of CRL 2688 made by \citet{cohen80}.

\subsection{\mbox{Na\,{\sc i}} and \mbox{K\,{\sc i}} resonance lines}

\begin{figure}
\begin{center}
\includegraphics[width=8cm]{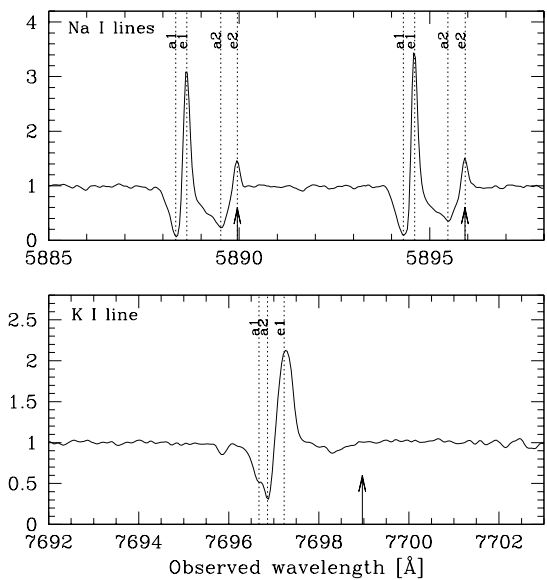}
\caption{\mbox{Na\,{\sc i}} and \mbox{K\,{\sc i}} lines. Emission and absorption components for which radial velocities of the line center have been measured are indicated by dotted lines. Their restframe wavelengths are indicated by arrows.}
\label{fig:na_k}
\end{center}
\end{figure}

Figure \ref{fig:na_k} shows portions of the 
spectrum around \mbox{Na\,{\sc i}} and \mbox{K\,{\sc i}} resonance lines. 
\mbox{Na\,{\sc i}} lines are expected to have contribution from 
stellar photosphere, circumstellar envelope and interstellar matter. 
However as we will indicate in Section \ref{sec:dib}, 
contribution from the interstellar absorption is likely 
small for this object. 
 The observed \mbox{Na\,{\sc i}} D1 and D2 lines show a complex 
structure 
showing multiple components. Each of the  
D1 and D2 line shows at least two emission components ('e1' and 'e2' 
in the top panel of Figure \ref{fig:na_k}). 
The stronger component ('e1') is blueshifted while the 
weaker component ('e2') is located close to the restframe wavelength.
The \mbox{K\,{\sc i}} lines of this star are likely 
of circumstellar origin since K would be mostly ionized in the 
photosphere.    
Similar to the \mbox{Na\,{\sc i}} lines, the \mbox{K\,{\sc i}} 
line shows both absorption and 
emission components ('a1','a2' and 'e1'). 
The absorption component, which is at 
the bluer side of the emission feature, appear to show at least 
two velocity components. 

Signatures of multiple components of \mbox{Na\,{\sc i}} and \mbox{K\,{\sc i}} 
absorption features have been reported in other post-AGB stars, 
like e.g. in HD 56126 \citep{bakker96,crawford00}. 
Based on a very high-resolution ($R\sim 900000$) observations 
of this object, 
\citet{crawford00} reported that the \mbox{K\,{\sc i}} 
absorption feature of this object 
shows multiple components, some of which have a counterpart 
in C$_{2}$ molecular features. 
They suggested that the individual components are associated 
with distinct 
shells of circumstellar matter around this object. 
Higher resolution spectroscopy is clearly needed to 
address the nature of the shape of \mbox{Na\,{\sc i}} and 
\mbox{K\,{\sc i}} lines 
in CRL 2688 by resolving their intrinsic profiles. 

\citet{klochkova04} reported that the \mbox{Na\,{\sc i}} 
emission component is likely unpolarized, suggesting that 
it is came directory from the observed lobe region.   
The presence of the emission components in both \mbox{Na\,{\sc i}} and 
\mbox{K\,{\sc i}} lines as seen in CRL 2688 
has also been observed in a number of other C-rich (and O-rich) 
post-AGB stars \citep{luna08}, 
although some of the similar class of objects like IRAS 06530-0123 
do not seem to show these emissions \citep{hrivnak03}.

\subsection{Diffuse interstellar bands}
\label{sec:dib}

Equivalent widths of diffuse interstellar bands (DIBs) have 
been calibrated to estimate the line-of-sight reddening ($E(B-V)$) \citep{friedman11}.  In the present spectrum of CRL2688, most of the reported 
strong DIBs are too weak or likely 
contaminated by neighboring spectral features. The well calibrated 
DIB centered at 5780.5 {\AA} may also be contaminated by photospheric 
absorption lines. Even though, it 
may be useful to estimate an upper limit of $E(B-V)$
toward the line-of-sight of CRL 2688.  
Note that a Doppler shift of the observed DIB is not consistent 
with that of the molecular absorption and emission lines, which
indicates the DIB is originated from interstellar matter rather
than from circumstellar envelope. Indeed, \citet{luna08} 
did not detect signatures of the diffuse 
bands of circumstellar origin 
for a representative sample of Galactic post-AGB stars.

The equivalent width of the feature near 5780.5 {\AA} is 67 m{\AA}. 
According to the calibration of \citet{friedman11}, this corresponds 
to $E(B-V)=0.1$, suggesting that the interstellar reddening is 
very low at the line-of-sight of this object. Although 
CRL 2688 is located at 
low Galactic latitude at $(l,b)=(80.166,-6.502)$, where 
integrated reddening is $E(B-V)=0.28$ according to \citet{schlegel98},
the low $E(B-V)$ value 
may be reasonable since the distance to this object was estimated to be 
only 420 pc \citep{ueta06}.  
On the other hand, photometric observation of \citet{ney75} suggests that the 
extinction toward this object in visual wavelengths is at least 
3 mag. If the interstellar reddening is very low as suggested by the 
present analysis, most of the extinction toward the CRL 2688 is due 
to the circumstellar matter and the dusty disk.

\section{Radial velocity}

Radial velocities from unblended metal absorption lines have been 
measured by fitting Gaussian to individual features. For the spectral 
lines containing both emission and absorption components, 
we have only used a core region of each feature to determine its 
central wavelength with the IRAF {\it splot} routine. 
In the following, all radial velocities are those corrected to 
helio-centric values ($V_{\sun}$).
Table \ref{tab:rv} summarizes heliocentric radial velocities obtained from the 
observed  spectral features.

The mean of radial velocities from unblended 
metal absorption lines is $-14.4\pm 0.6$ km s$^{-1}$, which represents 
the radial velocity of the central star. The standard deviation of the 
radial velocities from individual lines is $\sigma=2.7$ km s$^{-1}$. 
In the following, an error in a velocity measured from a single 
line is taken to be equal to this value. 
The $V_{\sun}$ measured from our spectrum 
is in good agreement with that obtained by \citet{klochkova00, klochkova04}  
using metal-lines in the spectrum, although the observed 
spectrum by \citet{klochkova00, klochkova04} is that of the northern lobe 
of CRL 2688, while the present study is based on the spectrum 
of the central star.

Emission components of \mbox{H\,$\alpha$}, \mbox{Na\,{\sc i}} ('e1' in Figure \ref{fig:na_k}) and \mbox{K\,{\sc i}} lines all
show similar radial velocity with a mean 
$-55$ km s$^{-1}$ with respect to their 
rest frame wavelength. 
This value is smaller in magnitude by more than 
$\sim 20$ km s$^{-1}$ from that 
obtained from the emission component of 
\mbox{H\,$\alpha$} by \citet{klochkova00}. 
As reported by \citet{klochkova00}, 
the emission components also 
displayed some variability in time. 
Therefore, the observed difference in the radial velocity 
may be attributed 
to both spatial and time variations possibly 
due to the circumstellar outflow.
The weaker emission components in the \mbox{Na\,{\sc i}} 
D1 and D2 lines ('e2' in Figure \ref{fig:na_k}) marginally agree with 
lab wavelengths. Therefore, it is likely that the 
 e2 components in the \mbox{Na\,{\sc i}} D1 and D2 lines 
are due to night sky emission and/or there may be some 
contribution from CRL 2688. 

The absorption component in the \mbox{H\,$\alpha$} line shows the 
radial velocity of $-10.6\pm 2.7$ km s$^{-1}$. This value 
is 3.8 km s$^{-1}$ smaller in magnitude than that from metal lines. 
This difference could be attributed to the presence of 
emission line in its violet wing, modifying the true 
central wavelength of the absorption feature. Similarly, 
\mbox{Na\,{\sc i}} D1 and D2 lines show a broad absorption 
feature ('a2' in Figure \ref{fig:na_k}) with a violet wing. 
The radial velocity of this feature is $-10.6$ and $-10.0$ km 
s$^{-1}$ for the D1 and D2 lines, which are similar to 
the absorption component of \mbox{H\,$\alpha$}. Another absorption 
components ('a1' in Figure \ref{fig:na_k}) of the \mbox{Na\,{\sc i} }
lines have radial velocity of $-69.8$ km s$^{-1}$. 
This value is close to the absorption component ('a2' in Figure \ref{fig:na_k})
 of the \mbox{K\,{\sc i}} line. 

The emission components of individual features in 
C$_{2}$ Phillips (3,0) and (2,0) bands show  
radial velocity of $-53.3\pm 0.9$ and $-51.7\pm 1.0$ km s$^{-1}$, 
respectively, that are marginally consistent with 
those of the \mbox{H\,$\alpha$}, \mbox{Na\,{\sc i}} and 
\mbox{K\,{\sc i}} lines. For the CN red system,
 the radial velocities for the different emission components are
difficult to measure because the emission features 
are weak and appeared to be affected by blending.

On the other hand, the absorption components of the 
C$_{2}$ Phillips bands 
 have a mean radial velocity of $-67$ km s$^{-1}$, 
which is more highly blue shifted
than the corresponding emission components, while   
the absorption components of the CN red system show 
a similar radial velocity of $-66$ km s$^{-1}$.

\begin{table}
 \centering
  \caption{Radial velocities}
  \label{tab:rv}
\begin{minipage}{6.5cm}
  \begin{tabular}{@{}lcc@{}}
  \hline
   Features & \multicolumn{2}{c}{$V_{\sun}$ (km s$^{-1}$)} \\ 
  & absorption& emission  \\ \hline
Metal lines & $-14.4\pm 0.6$ & ---  \\ 
\mbox{H\,$\alpha$}\footnote{Errors in $V_{\sun}$ measured from a single line 
are assumed to be 2.7 km s$^{-1}$, which is equal to the scatter ($\sigma$) 
in $V_{\sun}$ from the metal lines.} & $-10.6$ & $-55.0$  \\
NaI D1 & $-69.7$, $-10.6$ & $-54.8$, $12.7$ \\  
NaI D2 & $-69.9$, $-10.0$ &  $-55.1$, $12.3$ \\ 
KI & $-77.15$, $-69.67$\footnote{Deblending and fitting were not successfully made. The values for approximate local minima are given.} & $-55.2$  \\
C$_{2}$ Phillips (3,0) &$-67.2\pm 0.2$(3)\footnote{Number of lines used to measure the mean $V_{\sun}$ for each band.} &$-53.3\pm 0.9$(5)  \\ 
C$_{2}$ Phillips (2,0) &$-67.5\pm 0.6$(3) &$-51.7\pm 1.0$(5)  \\
CN red (3,0) & $-65.5\pm 0.2$(3) & --- \\
CN red (2,0) & $-66.8\pm 0.5$(4) & --- \\
\hline
\end{tabular}
\end{minipage}
\end{table}

\section{Abundance analysis}

The abundance analysis is performed 
using a LTE abundance analysis code as in 
\citet{aoki09} together with Kurucz NEWODF model atmosphere 
\citep{castelli03} using the EWs measured for the atomic 
absorption lines. In order to exclude the features 
with erroneous fitting, we restrict the lines 
used in the abundance analysis to those with a FWHM in a range 0.4 to 1.0, 
where typical FWHM of the metal absorption lines in our spectrum 
is 0.7 {\AA}. We also exclude very strong lines with  
$\log (EW/\lambda)\geq -4.4$. The measured EWs with
these criteria are used in the abundance analysis and  
are listed in Table \ref{tab:ew}.

\subsection{Atomic data}
We have adopted a linelist compiled from \citet{klochkova00} 
and \citet{reyniers04}. 
The adopted $\log gf$ values and measured equivalent widths 
are summarized in Table \ref{tab:ew}.

\begin{table}
\scriptsize
\centering
\caption{Equivalent widths}
\label{tab:ew}
\begin{minipage}{6.5cm}
\begin{tabular}{lcccccl}
\hline
Elem & $\lambda$ & $\log gf$ & $\chi$ & EW & Ref.\footnote{Reference for $\log gf$. R04: \citet{reyniers04}, K00: \citet{klochkova00}} \\
   & ({\AA}) & (dex) & (eV) & (m{\AA}) &  \\
\hline
   C  I &    4775.90 &  -2.19 &   7.49 &  136.29  & R04       \\
   C  I &    5039.06 &  -1.77 &   7.95 &   98.71  & R04       \\
   C  I &    6013.16 &  -1.16 &   8.65 &   94.61  & R04       \\
   C  I &    6014.83 &  -1.58 &   8.64 &   63.62  & R04       \\
   C  I &    7113.18 &  -0.93 &   8.65 &  100.95  & K00       \\
   N  I &    7468.31 &  -0.17 &  10.34 &  147.93  & R04       \\
   O  I &    6158.19 &  -0.31 &  10.74 &  119.67  & R04       \\
  Na  I &    5682.63 &  -0.67 &   2.10 &   43.70  & R04       \\
  Na  I &    5688.21 &  -0.37 &   2.10 &  117.00  & R04       \\
  Si II &    5056.02 &   0.31 &  10.03 &  172.29  & R04       \\
  Si II &    5978.93 &  -0.06 &  10.07 &   81.26  & K00       \\
   S  I &    6041.92 &  -1.00 &   7.87 &   21.87  & K00       \\
   S  I &    6743.58 &  -0.56 &   7.87 &   36.05  & R04       \\
   S  I &    6757.16 &  -0.20 &   7.87 &   65.03  & R04       \\
  Ca  I &    5857.46 &   0.24 &   2.93 &   72.66  & K00       \\
  Ca  I &    6439.07 &   0.39 &   2.53 &  111.29  & R04       \\
  Ca  I &    6462.57 &   0.31 &   2.52 &  102.07  & R04       \\
  Ca  I &    6717.69 &  -0.52 &   2.71 &   50.41  & K00       \\
  Sc II &    5640.99 &  -1.01 &   1.50 &  194.12  & K00       \\
  Sc II &    5667.15 &  -1.21 &   1.50 &  112.34  & R04       \\
  Sc II &    5669.04 &  -1.09 &   1.50 &  186.12  & K00       \\
  Sc II &    5684.20 &  -1.01 &   1.51 &  188.86  & K00       \\
  Ti II &    4798.52 &  -2.67 &   1.08 &  158.12  & R04       \\
  Ti II &    5418.80 &  -2.17 &   1.58 &  178.38  & K00       \\
  Cr II &    4812.35 &  -1.80 &   3.86 &  153.08  & R04       \\
  Cr II &    5246.77 &  -2.47 &   3.71 &  107.48  & R04       \\
  Cr II &    5249.43 &  -2.62 &   3.76 &   86.20  & K00       \\
  Cr II &    5305.87 &  -2.08 &   3.83 &  136.25  & R04       \\
  Cr II &    5308.43 &  -1.81 &   4.07 &  112.35  & R04       \\
  Cr II &    5334.87 &  -1.89 &   4.07 &  196.78  & R04       \\
  Cr II &    5407.60 &  -2.15 &   3.83 &  102.92  & R04       \\
  Cr II &    5420.92 &  -2.46 &   3.76 &  110.42  & R04       \\
  Mn  I &    4783.42 &   0.04 &   2.30 &   66.73  & R04       \\
  Fe  I &    4966.09 &  -0.84 &   3.33 &   62.76  & R04       \\
  Fe  I &    5049.82 &  -1.35 &   2.28 &   50.84  & R04       \\
  Fe  I &    5302.30 &  -0.88 &   3.28 &   79.74  & K00       \\
  Fe  I &    5364.87 &   0.23 &   4.45 &   67.96  & R04       \\
  Fe  I &    5367.47 &   0.44 &   4.42 &   84.23  & R04       \\
  Fe  I &    5393.17 &  -0.91 &   3.24 &   86.78  & K00       \\
  Fe  I &    5405.77 &  -1.84 &   0.99 &  163.20  & K00       \\
  Fe  I &    5434.52 &  -2.12 &   1.01 &   97.56  & R04       \\
  Fe  I &    5445.04 &   0.04 &   4.39 &   55.27  & R04       \\
  Fe  I &    5446.92 &  -1.93 &   0.99 &  162.08  & K00       \\
  Fe  I &    6024.06 &  -0.06 &   4.55 &   71.55  & R04       \\
  Fe  I &    6230.72 &  -1.28 &   2.56 &   28.13  & R04       \\
  Fe  I &    6400.00 &  -0.29 &   3.60 &   59.87  & R04       \\
  Fe II &    5991.37 &  -3.56 &   3.15 &  151.98  & R04       \\
  Fe II &    6084.10 &  -3.80 &   3.20 &  108.07  & R04       \\
  Fe II &    6416.92 &  -2.85 &   3.89 &  195.32  & K00       \\
  Fe II &    6432.68 &  -3.71 &   2.89 &  144.65  & R04       \\
  Fe II &    7711.71 &  -2.74 &   3.90 &  181.66  & K00       \\
  Ni  I &    5035.37 &   0.29 &   3.63 &   87.94  & R04       \\
  Zn  I &    4810.54 &  -0.17 &   4.08 &   28.38  & R04       \\
   Y II &    5289.81 &  -1.85 &   1.03 &   46.98  & R04       \\
   Y II &    5728.89 &  -1.12 &   1.84 &  168.35  & K00       \\
  Ba II &    5853.67 &  -1.00 &   0.60 &  228.35  & K00       \\
\hline
\end{tabular}
\end{minipage}
\end{table}

\subsection{Atmospheric parameters}

The effective temperature ($T_{\rm eff}$) of CRL 2688 was derived 
using the Fe I excitation equilibrium, in such a way that 
the trend of the Fe abundances from individual Fe I lines 
with their excitation potentials is minimized within 
$0.04$ dex eV$^{-1}$, which is the typical error of the slope. 
This yields $T_{\rm eff}=7250\pm400$ K. This value 
is higher than the previous estimate by \citet{klochkova00}, 
($T_{\rm eff}=6500$ K with a typical uncertainty of 200 K) 
but in good agreement with that obtained by 
\citet{bakker97} ($T_{\rm eff}=6900$ K) within the quoted error. 
The surface gravity ($\log g$) is estimated 
from the Fe I and Fe II ionization balance. 
The model atmosphere with $\log g=0.5$ dex, which is a lower limit of 
$\log g$ in the available model grids with $T_{\rm eff}>7000$ K, yields 
reasonable agreement between the Fe abundances based on 
the neutral and ionized species. 
Finally, the micro-turbulence velocity $\xi$ 
is estimated by minimizing the Fe I abundance versus 
equivalent widths correlation. This results in $\xi=6.5$ 
km s$^{-1}$ which agrees with that from \citet{klochkova00}.

\subsection{Abundances}

Adopting the atmospheric parameters estimated above, 
we have obtained abundances of C, N, O, Na, $\alpha$ elements, 
Fe-peak elements and neutron capture elements from 
the measured equivalent widths. 
For heavy neutron-capture 
elements, spectral synthesis have been applied, which 
gives upper limits in the abundances. 
In the following, we have taken an error as a line-to-line 
scatter divided by the square root of the number of lines 
(e.g. an error in the mean of the abundances from individual 
lines). When only one line is available for the abundance
estimate, then an error is taken to be equal to 0.25, which 
is the line-to-line scatter in abundances 
derived from individual Fe I lines.  

However, uncertainty from systematic errors may be 
much larger than the random errors from the line-to-line scatter. 
Systematic errors due to the adopted atmospheric parameters are checked by
changing the parameters by $\Delta T_{\rm eff}=\pm 400$ K, 
$\Delta\log g=+0.5$ dex (a model atmosphere with $\log g=0.0$ 
dex is not available) and  $\Delta\xi=\pm2.0$ km s$^{-1}$.  
 The results of this exercise and obtained [X/Fe] values
are summarized in Table \ref{tab:abu}.
Typically, changes in [X/Fe] values are $<0.3$ dex 
for the change in the $T_{\rm eff}$ and $<0.1$ dex 
for the change in the $\log g$, 
except for [N/Fe], [O/Fe] and [\mbox{Fe\,{\sc i}}/H]. 
 The [X/Fe] values change less than 0.25 dex for the
change in the $\xi$ by 2.0 km s$^{-1}$.

\begin{figure}
\begin{center}
\includegraphics[width=8cm]{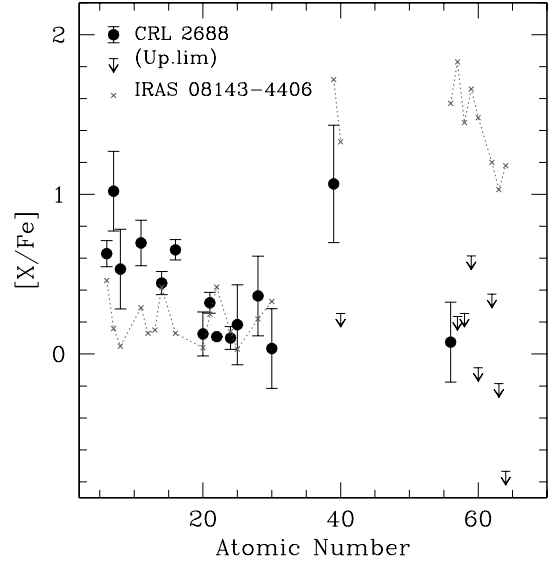}
\caption{Abundance pattern of CRL2688. Gray crosses and dotted lines show 
the abundance pattern of another carbon-rich post AGB star 
IRAS 08143-4406 from \citet{reyniers04}. }
\label{fig:abupattern}
\end{center}
\end{figure}

Figure \ref{fig:abupattern} shows the abundance ratios ([X/Fe])
as a function of the atomic number. 

\subsubsection{CNO and Na}
The abundance pattern of CRL2688 shows enhancement of 
C, N, O relative to Fe.
 The obtained abundance ratios, [C/Fe]$=0.63\pm 0.08$ and 
[O/Fe]$=0.53\pm 0.25$, 
are in good agreement with those derived by \citet{klochkova00}. 

The carbon abundance is estimated from the five lines with 
the line-to-line scatter of 0.18 dex. The [C/Fe] does
not significantly change by the change in $T_{\rm eff}$ as
shown in Table \ref{tab:abu}.  

The oxygen abundance is derived from the \mbox{O\,{\sc i}} 
line at 6158.2 {\AA}.  The abundance may be slightly overestimated 
since this line is blended with 
a weak \mbox{Fe\,{\sc i}} line at  6157.7 {\AA}. By synthesizing 
a spectrum around this line for this star, the contribution from the 
\mbox{Fe\,{\sc i}} line is estimated to be $\sim 4$ \%, 
which is much smaller than the other errors in [O/Fe].  
The derived abundance is also sensitive to the adopted temperature 
(Table \ref{tab:abu}). Furthermore, negative non-LTE correction 
of $0.1-0.2$ dex is expected for stars with a similar spectral type 
\citep{takeda98}. Therefore, an overabundance of [O/Fe] for this star 
is not yet clear in the present study.

The [N/Fe] ratio of $1.02\pm 0.25$ is significantly 
lower than that from \citet{klochkova00}, who derived [N/Fe]$=2.0$.  
Since as indicated in Table \ref{tab:abu}, the derived [N/Fe] abundance 
largely depends on 
the choice of parameters in the model atmosphere, 
especially in $T_{\rm eff}$, the difference from \citet{klochkova00} 
may be explained by systematic errors due to the atmospheric 
parameters as well as by difference in the measured EWs.  
Moreover, the N abundance from the \mbox{N\,{\sc i}} 7468.3 {\AA} 
line may suffer from non-LTE effects as suggested by \citet{takeda95}. 
The non-LTE analysis by \citet{takeda95} suggests that non-LTE 
corrections for the abundances derived from \mbox{N\,{\sc i}} 
lines around $\sim 8700$ {\AA} for F-type supergiant 
stars are $\sim -0.4$ dex. A similar non-LTE correction 
is expected for the \mbox{N\,{\sc i}} line used in this work.   
Taking into account a possible non-LTE correction of $-0.4$ dex, 
a modest [N/Fe] remains, although the non-LTE calculation 
is needed for a more reliable estimate.  

The Na abundance is estimated from two lines 
(5682.63 {\AA} and 5688.21 {\AA}). For these lines, 
\citet{takeda03} reported that non-LTE correction to 
the abundances of their sample 
stars is less than 0.2 dex, which is not significant within 
the precision of the present analysis.
The derived abundance ratio is [Na/Fe]$=0.70\pm 0.14$, which is 
in agreement with that derived by \citet{klochkova00}. 
This value is higher than that observed in most of the previously studied 
disk stars with similar metallicity, that generally show 
the solar ratio \citep{takeda03}. 

\subsubsection{Neutron-capture elements}

The Y abundance is estimated to be [Y/Fe]=$1.07\pm 0.37$ from two lines, 
where the large error comes from the line-to-line difference in 
derived abundances. This value is in 
agreement with that derived by \citet{klochkova00} and significantly higher  
than those observed in typical disk stars 
with similar metallicity \citep{edvardsson93}.   
Such enhanced [Y/Fe], however, is not unusual for other carbon-rich 
post-AGB stars \citep{reyniers04,reddy97}.

On the other hand, the abundance of the heavier 
s-process element Ba is not as highly enhanced as that of Y. 
We note that the Ba abundance was estimated from just one of the three 
lines detected in our spectrum 
because the other two lines seem to be affected by 
the emission in the violet wing (Figure \ref{fig:pcyg}).
The Ba line at 5853.67 {\AA} has an approximately symmetric profile, 
which we think it is more reliable.  
For the group of the heavy s-process elements,  
we could only obtain the upper limits of the abundances. 
Enhancements similar to [Y/Fe] are certainly not observed and remain 
comparable to or lower than the [Ba/Fe] ratio.
 
\begin{figure}
\begin{center}
\includegraphics[width=8cm]{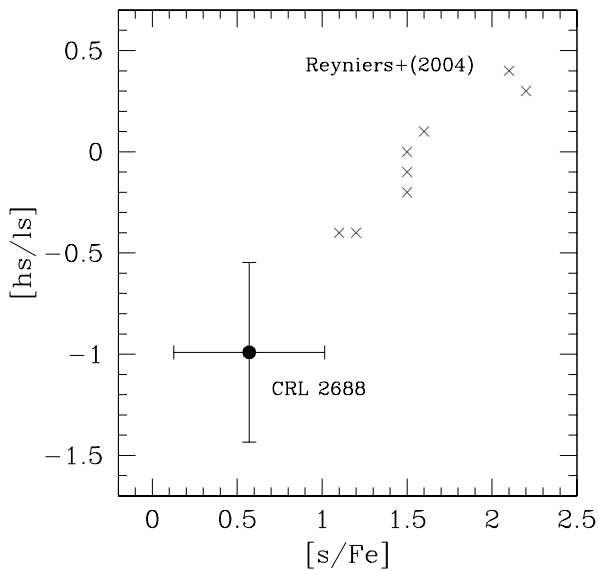}
\caption{Heavy s-process element to light s-process element ratios 
plotted against mean of s-process to Fe ratios. In this work (filled 
circle), 
[s/Fe] is taken as a mean of [Y/Fe] and [Ba/Fe], while [Ba/Y] ratio 
is used as a proxy of [hs/ls]. For the data obtained by \citet{reyniers04} 
(gray crosses), the definition from the original paper is used for both 
quantities. }
\label{fig:hs_ls}
\end{center}
\end{figure}

The ratio of heavy to light s-process elements 
is frequently used as a measure of the neutron exposure responsible for 
processing these elements. The observed [Ba/Y]$=-1.0\pm0.4$ is actually 
very low. This result has also been reported by the previous study of \citet{klochkova00}. 
In Figure \ref{fig:hs_ls}, we compared the abundance pattern of the s-process 
elements of CRL 2688 in this work 
with that of other post-AGB stars analyzed by \citet{reyniers04}. 
Here we use Ba and Y as a tracer for heavy-s and light-s 
process elements ([hs/ls]$\sim$[Ba/Y] and [s/Fe]$\sim$([Y/Fe]+[Ba/Fe])/2). 
In \citet{reyniers04}, the heavy-s process abundance is calculated as 
 the mean of the Ba, La, Nd and Sm abundances, while 
the light-s process abundance is the mean of the abundances 
derived from Y and Zr. Figure 
\ref{fig:hs_ls} shows that CRL 2688 has the lowest [hs/ls] compared to
the other objects. However, CRL 2688 appears to follow 
the same linear relation in [hs/ls]-[s/Fe].  

\subsubsection{Alpha and Fe-peak elements}

The $\alpha$ and Fe-peak elements are normal compared to the 
values observed in the Galactic disk stars with 
similar metallicity. In particular, the non-refractory element Zn 
abundance is estimated to be [Zn/Fe]$=0.04\pm 0.25$ dex, which 
is the standard value observed in disk stars with similar 
metallicity, suggesting 
that depletion of refractory elements due to condensation 
onto dust grains does not significantly contribute to the 
observed abundance pattern. Another non-refractory element, 
sulphur, shows an enhanced abundance ratio of [S/Fe]$=0.65\pm 0.06$ consistent 
with the previous analysis made by \citet{klochkova00}. This value is 
higher than typical values reported for disk 
dwarf/giant stars with similar metallicity studied by \citet{takeda11}, 
that generally show the solar ratio. Since the sulphur abundance is 
measured from relatively weak lines, a higher signal-to-noise 
ratio spectroscopy is needed to confirm the overabundance of [S/Fe].

\begin{table*}
\begin{minipage}{126mm}
  \caption{Abundance results}
  \label{tab:abu}
  \begin{tabular}{lccccccccc}
  \hline
   Elem &  [X/H] & [X/Fe] & $\sigma_{\rm lines}$\footnote{When only one line is 
available for the abundance estimate, then an error is taken to be equal to 0.25, which 
is the line-to-line scatter in abundances 
derived from individual Fe I lines. }&N & \multicolumn{2}{c}{$T_{\rm eff}$\footnote{Effect of errors in atmospheric parameters on [X/Fe] ratios. }}  & $\log g$ & \multicolumn{2}{c}{$\xi$}  \\
    &  (dex) & (dex) & (dex) & & $+400$ K & $-400$ K & $+0.5$ dex   & $+2.0$ km s$^{-1}$ & $-2.0$ km s$^{-1}$ \\
  \hline
 C  I &     0.31&     0.63&     0.08&  5&    -0.13&     0.20&     0.07&    -0.01&    -0.00\\
 N  I &     0.70&     1.02&     --- &  1&    -0.41&     0.52&     0.24&    -0.09&     0.16\\
 O  I &     0.21&     0.53&     --- &  1&    -0.42&     0.56&     0.26&    -0.06&     0.09\\
Na  I &     0.37&     0.69&     0.14&  2&    -0.08&     0.08&     0.00&     0.00&    -0.01\\
Si II &     0.11&     0.45&     0.07&  2&    -0.30&     0.33&     0.05&    -0.02&    -0.00\\
 S  I &     0.33&     0.65&     0.06&  3&    -0.11&     0.13&     0.02&     0.02&    -0.05\\
Ca  I &    -0.20&     0.12&     0.14&  4&    -0.05&     0.05&    -0.00&    -0.00&     0.00\\
Sc II &    -0.01&     0.32&     0.07&  4&     0.08&    -0.07&     0.01&     0.00&     0.01\\
Ti II &    -0.22&     0.11&     0.00&  2&     0.08&    -0.07&     0.01&     0.01&     0.00\\
Cr II &    -0.23&     0.10&     0.07&  8&    -0.02&     0.02&     0.01&     0.03&    -0.08\\
Mn  I &    -0.14&     0.18&     --- &  1&     0.02&    -0.01&     0.00&     0.01&    -0.03\\
Fe  I &    -0.32&     0.00&     0.07& 13&     0.41&    -0.44&    -0.19&    -0.04&     0.09\\
Fe II &    -0.34&     0.00&     0.04&  5&     0.25&    -0.18&     0.05&    -0.10&     0.24\\
Ni  I &     0.04&     0.36&     --- &  1&    -0.02&     0.03&     0.00&     0.00&    -0.00\\
Zn  I &    -0.29&     0.04&     --- &  1&     0.00&     0.00&     0.01&     0.03&    -0.07\\
 Y II &     0.73&     1.07&     0.37&  2&     0.08&    -0.08&     0.00&     0.05&    -0.10\\
Ba II &    -0.26&     0.07&     --- &  1&     0.13&    -0.11&     0.01&    -0.07&     0.21\\
\hline
\end{tabular}
\end{minipage}
\end{table*}

\section{Discussion}

\subsection{Circumstellar envelope}
The radial velocities of the molecular lines provide an
estimation of the velocity of the circumstellar outflow where
these lines are formed. 
The emission components of the C$_{2}$ Phillips bands  
have a mean radial velocity of $-38.1$ km s$^{-1}$ with 
respect to the photospheric atomic absorption lines, while 
the emission components of the H$\alpha$, \mbox{Na\,{\sc i}} (e1) 
and \mbox{K\,{\sc i}} lines have a similar
value of $-41$ km s$^{-1}$.
  For the absorption components of the C$_{2}$ Phillips bands, 
this value is $-53.0$ km s$^{-1}$, which is in good 
agreement with the value of $-51.8$ km s$^{-1}$ derived 
from the CN red system.
Existence of several velocity components of circumstellar outflows 
in a range 19-100 km s$^{-1}$ 
have been identified based on CO and $^{13}$CO observations at
radio wavelengths \citep{young92,yamamura96}.

In the circumstellar environment, these molecules are thought to  form 
from more complex molecules like C$_{2}$H$_{2}$ 
and HCN that are photodissociated by 
interstellar and/or stellar radiation 
fields. Therefore, it is suggested that these molecules are confined 
to a thin circumstellar shell whose inner and outer radius are 
determined by the penetration/shields of photons from 
the radiation fields \citep{bakker97}. The presence of 
both absorption and emission components in C$_{2}$ Phillips bands 
suggests complex 
geometric structure of this circumstellar shell in which
C$_{2}$ and CN can exist.  Furthermore, the 
radial velocity differences between these components 
indicate that the regions where the absorption components of 
the C$_{2}$ and CN bands arise are expanding 
faster than those for the emission components.

\subsection{Abundance pattern}

 The enhanced C, N and Na abundances seen in CRL 2688
are indicative of dredge-up and mixing of 
H and He burning products during the AGB phase of evolution.
In this process, freshly synthesized s-process 
elements are also expected to be dredged-up, giving 
rise to overabundance in s-process elements in the photosphere. 
Indeed a class of carbon-enriched post-AGB stars are 
known to be enhanced with s-process elements \citep{hrivnak03,reyniers04}. 
However, not all of them 
show this enhancement and some stars show depression in 
the s-process elements instead.
The origin of this diversity in s-process abundance patterns is
not well understood \citep[e.g.][]{kappeler11}. Theoretical models 
for the generation of s-process elements during the AGB evolution 
depends strongly on the parametrization of unknown factors
such as a $^{13}$C pocket, which  may control neutron 
exposure \citep{busso99}. \citet{busso99} also suggested that 
the s-process efficiency depends on progenitor metallicity.
    
\citet{reyniers04} found a strong correlation 
of [hs/ls] index with the overall s-process elemental 
abundance ([s/Fe]) for their sample of carbon-rich post-AGB 
stars. They have suggested that the observed variation in 
the s-process elemental abundance patterns can partly be explained by 
AGB models with varying strength of the $^{13}$C pocket. 
The observed low [Ba/Y], treated as a proxy of 
[hs/ls], appears to follow the suggested [hs/ls]-[s/Fe] relation 
but with the lowest estimate of [hs/ls], which is 
lower than typical theoretical predictions \citep{kappeler11}. 
Alternatively, a neutron source of $^{22}$Ne($\alpha$,n)$^{25}$Mg, 
which would operate in rather massive AGB stars, might be 
responsible for the observed abundance pattern. If this is the case,
the progenitor may be a star with more than 3 or 4 solar masses.
However this would be in contradiction with the 
observed C-rich nature of the circumstellar envelope of CRL 2688.  
Note that the abundance result suggests that 
C/O$<$1 from the photospheric absorption lines. However, 
the oxygen lines may be influenced by non-LTE effects. If we apply
non-LTE correction to oxygen abundance the  C/O ratio my be 1.0 or more.

Since only Y and Ba have reliably been estimated in 
the present work, the conclusion about the very low [hs/ls] 
remains to be confirmed. Further accurate 
abundance study with much higher resolution
and high signal to ratio spectrum may enable us to 
understand nucleosynthesis in the progenitor star.

\section{Conclusion}
We have presented the analysis of a high-resolution optical 
spectrum of CRL 2688. 
Based on the one-dimensional LTE abundance analysis, we have 
confirmed that the central star has a spectral type of 
a F-type supergiant with 
[Fe/H]$=-0.3\pm 0.1$. Molecular absorptions from 
C$_{2}$ Phillips, C$_{2}$ Swan bands 
and CN red system, that are presumably of circumstellar origin 
 have been identified, some of which 
appear in emission. In particular, C$_{2}$ Phillips band 
show presence of emission components with the helio-centric 
radial velocity offsetted from the photospheric absorption 
lines by $-38$ km s$^{-1}$, which is consistent with 
emission components seen also in the H$\alpha$ line, \mbox{Na\,{\sc i}} and 
\mbox{K\,{\sc i}} resonance lines ($-41$ km s$^{-1}$).
The absorption components of these lines have a 
mean radial velocity of $-52$ km 
s$^{-1}$ with respect to the photospheric absorption lines, 
and are more highly blueshifted
than the corresponding emission components.

The abundance pattern of CRL2688 is characterized by 
the enhancement of C, N, Na and Y. However, unlike other known
C-rich post AGB stars, this object shows a very low [Ba/Y] ratio. 
Since only abundances of Ba and Y have been reliably measured 
as a tracer of 
heavy and light s-process elements, respectively, 
in the present work, the conclusion about the peculiar [hs/ls] 
ratio observed in CRL 2688 remains to be confirmed with 
measurements of more 
heavy and light s-process elements.

Higher resolution and higher signal-to-noise ratio 
spectroscopy is required  
for more comprehensive understanding 
of the origin of the each spectral feature and the 
detailed abundance pattern.

\section*{Acknowledgments}

MP is thankful to Prof. Shoken Miyama  for his kind support, encouragement and hospitality. D.A.G.H. and A.M. also acknowledge support provided  
by the Spanish Ministry of Economy and Competitiveness under grant  
AYA2011-27754.

This publication is based on observations made on the island of La Palma
with the William Herschel Telescope (WHT). The WHT is operated by the  
Isaac Newton Group (ING) and is located in the Spanish Observatorio of  
the Roque de Los Muchachos of the Instituto de Astrofisica de Canarias.

\appendix

\label{lastpage}

\end{document}